# Use of coated silicon field emitters as neutralisers for fundamental physics space missions

K.L. Aplin[1] (a), B.J. Kent[2] (a), C.M. Collingwood[3] (a,b), L. Wang (c), R. Stevens(c), S.E. Huq(c) and A. Malik(c)

(a) *RAL Space, Rutherford Appleton Laboratory, Chilton, Didcot, Oxon, OX11 0QX UK*
(b) *Department of Physics, University of Surrey, Guildford, Surrey, GU2 7XH, UK*
(c) *Micro and Nanotechnology Centre Rutherford Appleton Laboratory, Chilton, Didcot, Oxon, OX11 0QX UK*

**Abstract**

Spacecraft neutralisers are required as part of the ion propulsion system for accurate station keeping in fundamental physics missions. This paper describes the use of thin layers of insulating materials as coatings for the gated silicon field emitter array structure used in a spacecraft neutraliser. These thin coatings are postulated to reduce power consumption and reduce overheating. The power consumption and lifetime of aluminium nitride and amorphous hydrogenated diamond-like carbon coatings have been tested by current-voltage and endurance tests. Diamond-like carbon coatings were promising, performing better in endurance tests than uncoated samples, but further work is required to characterise the coating's physical properties and its effects on field emission. The thermal conductivity of the coating material had little effect on measured sample lifetimes. Aluminium nitride had reduced power consumption compared to diamond-like carbon coated and uncoated samples. A thin (~5 nm) layer of aluminium nitride was found to be optimal, meeting European Space Agency specifications for the neutraliser engineering model.

## 1  Introduction

Tests of fundamental physics in space use the ultra-quiet environment of a free falling spacecraft. A satellite in free fall experiences only differential gravitational forces; the spacecraft is also remote from sources of gravitational disturbance such as large moving masses. However non-gravitational disturbances, like drag from the residual atmosphere in low Earth orbit, or solar radiation pressure during heliospheric missions, can still influence measurements at the highest precision. In order to achieve the pico-gravity environment required for fundamental physics missions a disturbance compensation system is used, often described as drag free control.

In drag free systems, station keeping is controlled by micro-propulsion, providing thrust in the micro- to milli-Newton range. In the European context, the preferred

---

[1] Now at University of Oxford, Physics Department, Denys Wilkinson Building, Keble Road, Oxford, OX1 3RH UK (k.aplin1@physics.ox.ac.uk)
[2] Now retired
[3] Now at School of Engineering Sciences, University of Southampton, University Road, Southampton SO17 1BJ UK

micro thrusters are field effect electric propulsion devices or FEEPs. FEEPs use field emission to extract ions from a liquid metal, such as caesium or indium, which are then ejected at high velocity (e.g. Tajmar *et al*, 2004). The ion emission causes the spacecraft to quickly become charged to high negative potentials which reduces thrust efficiency, eventually to zero, and may result in damage to spacecraft systems by the return ion flux (Aplin and Tarakanov, 2004). All ion thruster systems therefore need a neutraliser.

For the Lisa Pathfinder mission, for which this system was intended, a single neutraliser serving a cluster of 4 FEEP thrusters minimises system power consumption and reduces complexity. The neutraliser must provide an electron current at least equal to the maximum current from four FEEP thrusters; for LISA Pathfinder this was 6mA. Power allocation for the LISA Pathfinder neutraliser is 0.2W/mA, which dictates the use of a cold emission mechanism such as field emission as the electron source, implying a maximum voltage of 200V at 6mA. The combination of these requirements, and an additional specification to keep the neutraliser mass below 100g, effectively restricts the production technology to micro-fabrication. This resulted in the choice of gated silicon tip field emitter technology for the neutraliser system, due to its high current density per unit mass and low power operation (Aplin *et al*, 2004). (Subsequent work has shown that carbon nanotubes are a more promising choice (Aplin *et al*, 2009), but field emitter tips were the more mature technology at the start of this research programme). This paper discusses the development of one aspect of the neutraliser system: the use of a coating to protect the neutraliser structure and improve performance. In the next section, the special requirements for using field emitters in fundamental physics space missions will be briefly described. Section 3 summarises the field emitter fabrication and apparatus used in the test programme. In Section 4, the basis for using coatings to improve field emitter performance is introduced, and the tests of coated field emitter arrays discussed in Section 5.

## 2   Design and operating conditions for field emitters for use in space

Much previous research into field emitters has focused on developing them for commercial applications such as flat screen displays. Whilst field emitters can generally perform well in the laboratory, some special considerations are needed when developing a device intended for use in the harsher environment of space. Fundamental physics missions will require the micro-propulsion system to be in almost continuous operation for the duration of the scientific data collection period. For LISA Pathfinder this is many months and thus a lifetime exceeding 6000 hours was demonstrated (Aplin *et al,* 2004; Kent *et al,* 2005).

Secondly, the neutraliser must be able to withstand operation at the exterior of a spacecraft, including the possibility of low fluxes of neutral metal atoms, or clusters of atoms, produced by charge exchange processes from the thruster. A build-up of metallic particles between the tip and gate could short out individual emitters, or in a worst case scenario, the entire neutraliser. The emission of electrons also generates ions in the region between the tip and gate, which may bombard the silicon tips. In higher-pressure environments such as low earth orbit there are also relatively high concentrations of atomic oxygen, which could attack the neutraliser structure.

This paper focuses on the use of a thin coating of insulating material to protect the silicon tip electron emitters against the risk of electrical short circuit. Coating silicon tips was initially suggested to overcome the perceived shortcomings of silicon, particularly mechanical fragility and low thermal conductivity, leading to the possibility of overheating. Attempts have previously been made to improve the robustness of the Si emitter tips by coating them with materials such as diamond like carbon (DLC), an artificially-produced form of amorphous carbon with a varying $sp^2/sp^3$ bond ratio (*e.g.* Xu *et al*, 2001) or aluminium nitride (AlN) (Zhirnov *et al*, 1997).

## 3   Overview of field emitter fabrication and test programme

The silicon field emitter arrays are fabricated using techniques described by Wang *et al* (2004, 2006). The complete neutraliser will utilise 66 2 x 7mm die, each containing 20 arrays, with each array holding 765 tips emitting ~7nA each, to make the total current 6mA. In the test programme described in this paper, individual chips were prepared, each containing 20 arrays on a 2 x 7mm die mounted on a dual in line package. The operating principle that if a voltage is applied to the chromium gate, which is electrically insulated from the grounded tip structure, the small dimensions and sharp point of the tip result in very high electric fields at the tip apex, where electrons are emitted by quantum tunnelling (e.g. Brodie and Spindt, 1992). A scanning electron micrograph of a typical field emitter array is shown in Figure 1.

The physics of field emission is described by the Fowler-Nordheim equation, and from it an exponential relationship can be derived between the voltage applied to the field emitters *V,* and the emitted current *I* (e.g. Brodie and Spindt, 1992). It can also be shown that if $ln(I/V^2)$ is plotted against $1/V$ then a linear relationship can be expected if field emission is occurring. Plotting the current-voltage response of field emitter arrays is a good test for field emission, and the current emitted at a particular voltage also indicates the power consumption of the device.

The laboratory test set up, described in detail in Aplin *et al* (2004), was therefore devised principally to conduct current-voltage tests of field emitter samples. In space the electrons are attracted to the positive ions emitted from the thruster, but in the laboratory test scenario a Faraday cup held at 300V is used to attract the electrons; the current to it was monitored using a Keithley 487 picoammeter/voltage source. The current emitted by the silicon tips and losses to the gate were measured using two more Keithley 487 instruments. Computer-controlled procedures were used to automatically increase and decrease the gate voltage and write data to an ascii file. One important aspect of field emitter behaviour is the need to "condition" arrays once they have been exposed to the ambient laboratory environment. This is the need to repeat I-V curve measurements until the results no longer show hysteresis effects, related to removal of surface impurities from the field emitter structure (Collingwood, 2004).

The same apparatus was used to measure the lifetime of samples running at a constant tip emission current, achieved by adjusting the gate voltage in software. Each array on the test chip needs to emit 5μA for the completed neutraliser to produce the 6mA needed to equal the ion beam current. Predefined criteria based on the mission specifications were chosen to define "failure": if the voltage required > 200V, or > 20% of the emitted current was lost to the gate electrode (Aplin *et al*, 2004). The nominal emission current for the arrays was defined as 6 μA, to allow for a 20% loss to the gate.

**4     Coating field emitter arrays**

The theoretical basis for improving silicon field emitter performance by adding coatings of wide band gap materials (WBGM) on field emitter tips was presented by Zhirnov *et al* (1997). If a field emitter tip is coated with a dielectric layer of a WBGM such as $CaF_2$, AlN or DLC, the electric field penetrates into and through the layer and is concentrated at the conductive tip beneath the dielectric. Electrons are assumed to either quantum tunnel from the semiconductor into the conduction band of the WBGM, where they are transported through the conduction band to the surface, or they tunnel into localised states in the WBGM and then move between these states to the surface. WBGM with negative electron affinity (NEA) surfaces, such as AlN and DLC, are often preferred for coatings, as the conduction band at the surface of such a WBGM may be pinned below the vacuum level, allowing the electrons to escape directly into a vacuum. Coating with NEA materials may therefore enhance emission, by reducing the effective work function.

There are additional, simple reasons why adding a coating to silicon field emitter arrays may be particularly beneficial in space applications. The coating can act as a protective layer against adsorption and ion bombardment. Adding a layer of material with a different thermal conductivity to silicon may also protect the tip against failure by melting. The enhanced performance expected from coated tips is also a strong motivation for using a coating in space, when efficiency and minimising power consumption are important.

Experimental evidence exists in the literature that both AlN and DLC coatings can improve the performance of silicon field emitter arrays. For example, Zhirnov *et al* (1997) reported that coating a silicon tip with AlN reduced the operating voltage by 33% and increased the current output. Jung *et al* (2000) also found that a DLC coating decreased the effective work function and reduced the operating voltage of the field emitter array. Both AlN and DLC coatings have been evaluated as part of our test programme, and Figure 2 shows the basic tip structure and the coating layer. AlN coatings were applied at the end of the tip fabrication process, using magnetron sputtering to deposit a layer of aluminium nitride over the whole array structure (Wang *et al*, 2004, 2006). Figure 3 shows micrographs of some of the tips tested.

DLC coatings were also applied at the end of the fabrication procedure, using plasma assisted chemical vapour deposition (PACVD) (e.g. Christiansen *et al*, 1996). The PACVD technique produces a hydrogenated DLC layer using a radio frequency plasma in a mixture of hydrogen and acetylene ($C_2H_2$) (Pierson, 1993). Unlike diamond and graphite, the physical structure, composition and properties of DLC are highly dependent on the $sp^2/sp^3$ bond ratio and the fraction of hydrogen present, which are in turn affected by the details of the processing technique. This can be an advantage as the fabrication technique can be tailored to achieve the desired properties (Pierson, 1993). However, the small quantity of samples used in this study did not justify the application of solid-state analysis techniques required to characterise them in detail.

**5     Testing coated arrays**

The availability of in-house AlN coating technology meant that it was possible to coat more field emitter test chips, and carry out more detailed tests on AlN than DLC coated samples. For the AlN coating, a calibration procedure was defined where blank silicon die were coated, and the thickness measured with a refractive index technique. The DLC coating thickness was estimated in the same way.

Standard tests, carried out on both AlN and DLC coated samples, were:
1. current-voltage curves: used to determine emission current, and projected power consumption, discussed in section 5.1
2. endurance tests: running the samples at constant current until they failed pre-defined performance criteria, defined in Section 3. The effects of different coatings on sample lifetime are described in section 5.2.

The effect of different thicknesses of AlN coating was also investigated by building up the thickness of the layer slowly on one sample and carrying out performance tests at each stage, with the results presented in section 5.4.

*5.1 Current-voltage tests*

29 AlN coated test arrays were subjected to current-voltage tests, compared to 6 DLC coated arrays. Before the arrays were conditioned, the DLC coated samples required considerably higher initiation threshold voltages, typically ~150V, than arrays coated with a similar thickness of AlN and uncoated arrays, which usually began to field emit at ~100V and 90V respectively. The typical I-V and Fowler-Nordheim characteristics of an AlN coated array and a DLC coated array of similar thickness are compared with the an uncoated array in Figure 4.

After conditioning, the DLC coated arrays generally operated at significantly higher voltages than the uncoated and AlN coated arrays, which is inconsistent with predictions that DLC coatings could improve emission efficiency (Xu *et al*, 2001). From the I-V curves, the average voltage required for a conditioned array to emit the nominal current of 6 μA (defined in section 2) was found to be 81 ± 2V for the AlN coated arrays and 93 ± 3V for the DLC coated arrays. Figure 5 shows the leakage to the gate electrode for similar thickness coatings of AlN and DLC. Both the current losses and voltages required are lower for the AlN coated sample, indicating it is the more power-efficient coating.

*5.2 Endurance tests*

Lifetimes of up to 6000 hours have already been demonstrated for AlN coated arrays (Aplin *et al*, 2004), indicating that AlN coated field emitter arrays can meet the ESA specification for the LISA Pathfinder satellite. This section therefore focuses on the typical characteristics of endurance tests on DLC coated arrays, compared to uncoated arrays. Figure 6 shows the results from a typical endurance test of two DLC coated arrays wired together. The sample was run at a constant emission current of 12 μA (Figure 6c), and failed when the gate current (Figure 6b) exceeded the limit of 20% of the tip current after ~40 hours. Figure 7 shows an endurance test run on an uncoated sample that suffered an abrupt failure after 23 hours.

There are differences between the uncoated and DLC coated samples. Firstly, the DLC current recorded at the collector (Figure 6a), is noisier at the start of the test than the uncoated trace. Collingwood (2004) suggested that the noise is caused by surface impurities due to inadequate conditioning, which are subsequently removed by heating during the endurance test (thermal effects are discussed further in section 5.3). The conditioning procedure was initially developed for uncoated tips, and appears to be adequate for AlN coated tips. This suggests that both uncoated and AlN coated tips may have similar surface properties, which are not the same as the DLC coated tips, though detailed surface analysis techniques would be required to check this. Secondly, as was indicated in section 5.2, the operating voltage of the DLC coated tips (Figure 6d) is higher than the uncoated tips (Figure 7d) and a greater voltage variation is required to hold the tips at constant current compared to the uncoated sample. Again, this may be related to inadequate conditioning at the start of the test. After 10 hours however, the gate voltage variability is still greater than for the uncoated tips, possibly suggesting the surface properties of the DLC tips vary more rapidly than the uncoated tips.

Some of the differences between Figures 6 and 7Figure  are likely to be related to the mechanism ultimately causing array failure, rather than coating effects. Using the failure classifications in Aplin *et al* (2004), the DLC coated array (Figure 6) failed gradually, and the uncoated array (Figure 7) failed abruptly. Assuming the increased variability in emitted current in the first ten hours for the DLC coated sample is due to inadequate conditioning, the emitted current (Figure 6c) shows an initially similar level of variability, ±16%, compared to the uncoated tips (Figure 7c). During the test, the variability of the DLC coated tip emission drops. The average current emitted by each tip increases non-linearly with a decreasing number of tips emitting (Shaw and Itoh, 2001). If all tips are assumed to have similar variability, the observed decrease in noise is consistent with fewer tips emitting. (The software feedback loop employed to control the emitted current also modulates the frequency response of the emission (Aplin et al, 2006)). This reduction in variability with time is almost certainly a symptom of the gradual deterioration of the array, as is the trend in gate voltage in Figure 6d. More endurance tests on different coated arrays would be required to determine how the coating influences failure type, as the array failure statistics from the relatively small number of samples tested are inconclusive.

*5.3   Thermal effects*

As mentioned in Section 4, the use of a layer of material of different thermal conductivity on silicon field emitter arrays may increase their lifetime. The proposed mechanism is that a coating of increased thermal conductivity can conduct heat away from the tip apex and make it less likely to melt at high current densities. The thermal conductivity of 2-4Ωm n-type Si has been estimated, corresponding to 1-3 x $10^{15}$ atoms/cc of phosphorous dopant. Thermal conductivities of Si, AlN and DLC are shown in Table 1. The table also indicates the number of samples suspected to have failed by overheating during endurance tests (described in Section 3 and in Aplin *et al*, (2004)). The number of thermal failures does not appear to be related to the thermal conductivity. DLC has the highest thermal conductivity, but had the highest fraction of assumed thermal failures. AlN had the lowest number of thermal failures, yet it has a lower thermal conductivity than silicon. Possible explanations for this are that the thermal conductivity of DLC is lower than the estimated value, or that the failure mechanism classifications are inadequate. However, thermal models of silicon field emitter tips, (Aplin *et al,* 2004), run for a wide range of silicon thermal conductivities, indicated that maximum tip temperatures, calculated for 6 µA passing through a single tip, were ~500°C. Thin surface coatings had a negligible effect when included in the thermal model, therefore it is possible that the number of abrupt failures is not caused by thermal effects (Shaw and Itoh, 2001).

*5.4 Effect of changing AlN coating thickness*

Two uncoated central wafer dies, 275 and 276, were characterised electrically by I-V tests and with an optical microscope. These tests indicated that the samples were similar in geometry and electrical performance. In an experiment to determine the optimum thickness of AlN, both dies were then sequentially coated with several layers of AlN. The range of AlN thickness investigated was ~5-20nm, estimated from a calibration test on a blank silicon wafer. 275 was initially coated with ~5nm AlN, and 276 with ~10nm, followed by subsequent ~5nm layers to compare the effect of increasing the coating thickness. I-V tests were carried out on both die between each coating stage.

Figure 8 shows the effect of adding successive layers of AlN to one sample, 275 (the second sample, 276 responded similarly). It is clear the presence of a thin (4.15nm) layer improves the performance of the sample by reducing the operating voltage and increasing the emission current. The absolute magnitude of the leakage current to the gate is the same as, or less than for the uncoated sample. The effective neutralisation current, i.e. the current leaving the array structure = (emission current - gate current) is therefore higher for 4.15nm coated than uncoated tips. Thicker coatings perform less well than the uncoated sample: both the emission current decreases, and the operating voltage increases with increasing thickness. The fraction of current returning to the gate electrode is high for the thickest (13.75nm) AlN coating and some resistive leakage current has developed which is apparent at low voltages, before field emission starts. The origin of this low voltage leakage current is unclear. Results from both samples are summarised in Table 2.

One explanation for the deterioration in performance with thicker coatings is that impurities could have been added between successive coatings; however, this seems unlikely, as the coatings took place in a class 100 clean room and the samples were transported in clean containers to the test facility. Additionally, the radio frequency plasma used in the coating process (Wang et al, 2004) is likely to have driven off any surface impurity layers, and the conditioning procedures also appeared to effectively stabilise the switch on voltage and emission characteristics, implying removal of surface impurities.

## 6 Conclusions

AlN and DLC-coated samples of silicon field emitter arrays have been compared, and AlN is preferred, on the basis of the samples tested. DLC coated samples had higher power consumption, running at voltages typically 30% higher than a similar thickness of AlN coating. Endurance tests showed that DLC coated samples lasted on average longer than uncoated samples but less long than AlN coated samples. In endurance tests, DLC had more variable emission characteristics than other samples, which is unexpected given that application of a DLC coating often makes a sample smoother than it was originally (Pierson, 1993). Further investigation is required to physically characterise the DLC samples tested in this paper. DLC coatings appear promising, but optimisation experiments similar to those carried out on AlN coated samples in Section 5.4 are required as well as analysis to characterise the physical properties of the layer, such as accurate thickness determination.

It has been suggested that coatings of higher thermal conductivity than silicon, such as DLC can reduce the incidence of tip melting. Results presented in this paper are inconsistent with this, as AlN coated tips, which have a lower thermal conductivity than silicon, survived for longer than both uncoated and DLC coated tips, with lower rates of thermal failure. Thermal modelling studies indicated that a thin AlN coating had minimal effect on field emitter temperatures, and it is possible that the coatings used are simply too thin to have any appreciable thermal effect. Surface effects may also play a role, but microscopy able to resolve surface morphology would be required to examine silicon tips and their coatings in detail.

Results of both current-voltage (Section 5.1) and endurance tests (Section 5.2; also see Aplin *et al* (2004) and Kent *et al* (2005)) on AlN coated silicon field emitter arrays showed that an AlN coating could meet the ESA power and reliability criteria for a neutraliser for fundamental physics missions. Experiments layering samples, selected for their consistent response and similarity to each other, with increasing thicknesses of AlN were described in Section 5.4. They indicated that a thin coating is optimal and enhances emission by reducing operating voltage and increasing current output, presumably by reducing the effective workfunction of the sample as discussed in Section 4. However thicker layers of AlN were less desirable, increasing operating voltage, and appearing to show some resistive leakage at voltages lower than the levels required for field emission. The AlN coating thickness for the FEEP neutraliser engineering model has been chosen based on the results of these experiments. Subsequent tests carried out on the final device at the European Space Agency were successful (Frigot *et al*, 2007), although ultimately, technologies that are simpler to fabricate, such as carbon nanotubes, may be more promising for use in space (Aplin *et al*, 2009).

**Acknowledgements**

This research was funded by the European Space Agency. K.J. Fereday carried out the thermal modelling work and helped to source some of the thermal reference data. We are grateful to the Manufacturing and Engineering Systems Department, Brunel University for applying the diamond-like-carbon coatings.

| Material | Estimated thermal conductivity (Wm$^{-1}$K$^{-1}$) at 300K | Number of abrupt failures in endurance tests | Reference |
|---|---|---|---|
| n-type silicon | 120-300 | 4 of 11 | Brinson and Dunstant (1970), Anon (1998) |
| Diamond-like carbon | 400-1000 | 2 of 3 | Pierson (1993) |
| Aluminium nitride | 20 | 1 of 3 | Shackleford (2001) |

Table 1 Table of estimated thermal conductivities compared to number of abrupt failures, thought to be due to thermal effects, in endurance tests. References for the thermal conductivity estimates are also given.

| Sample | Target AlN Coating Thickness / nm | Starting voltage for field emission / V | Gate voltage required for 5 µA tip current / V | Emitted tip current at 70V / µA | % emitted current recorded at gate |
|---|---|---|---|---|---|
| 275 | 0 | 58 | 66 | 11 | 12 |
| 276 | 0 | 59 | 75 | 1.7 | 12 |
| 275 | 5 | 53 | 62 | 12 | 4 |
| 276 | 10 | 65 | 87 | 0.5 | 25 |
| 275 | 10 (5+5) | 57 | 71 | 3 | 12 |
| 276 | 15 (10+5) | 65 | 84 | 0.5 | 33 |
| 275 | 15 (5+5+5) | 65 | 84 | 0.6 | 32 |
| 276 | 20 (10+5+5) | 65 | 90 | 0.7 | 30 |

Table 2 Summary of results from current-voltage tests on successive layers of aluminium nitride coating on two samples.

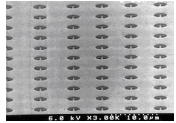

Figure 1

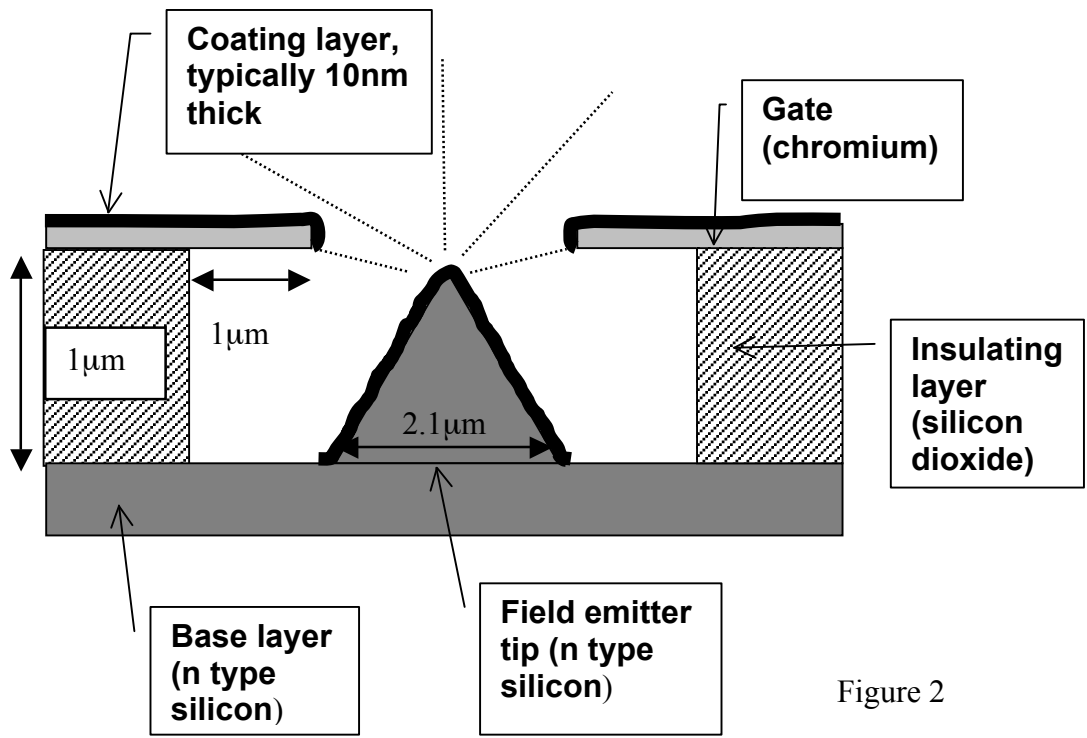

Figure 2

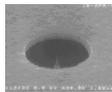

Figure 3(a)

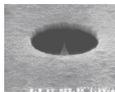

Figure 3(b)

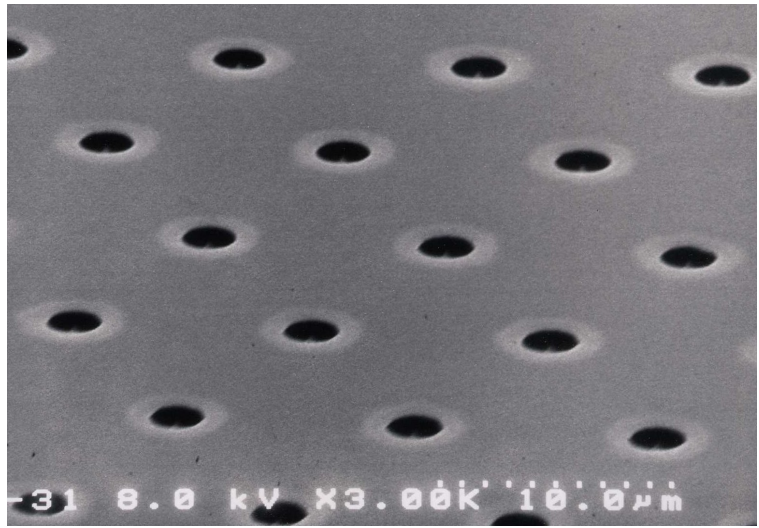

Figure 3 (c)

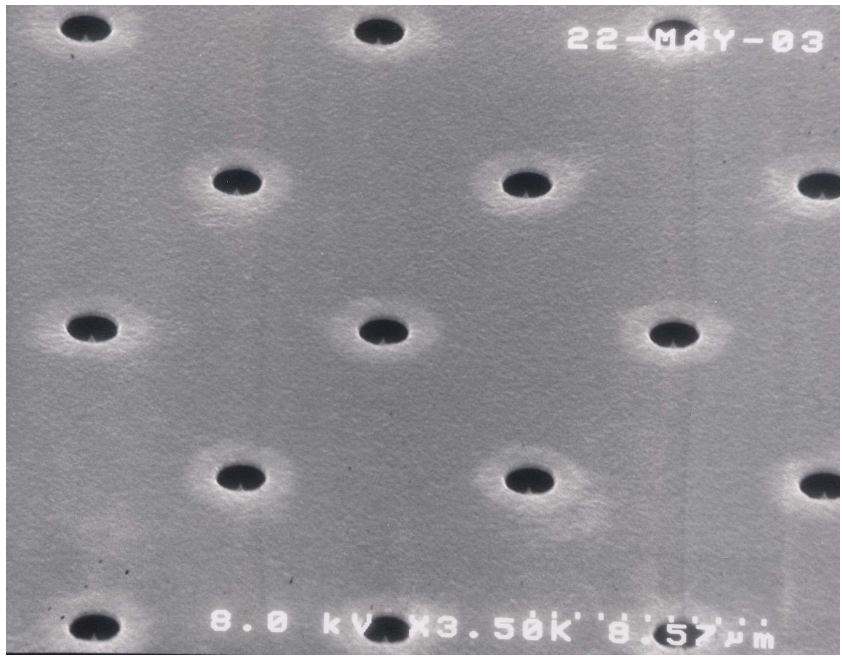

Figure 3(d)

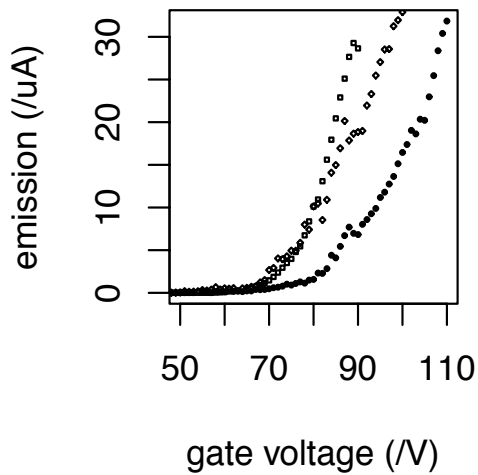 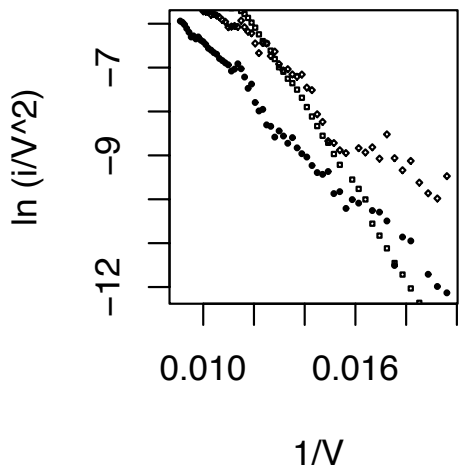

Figure 4

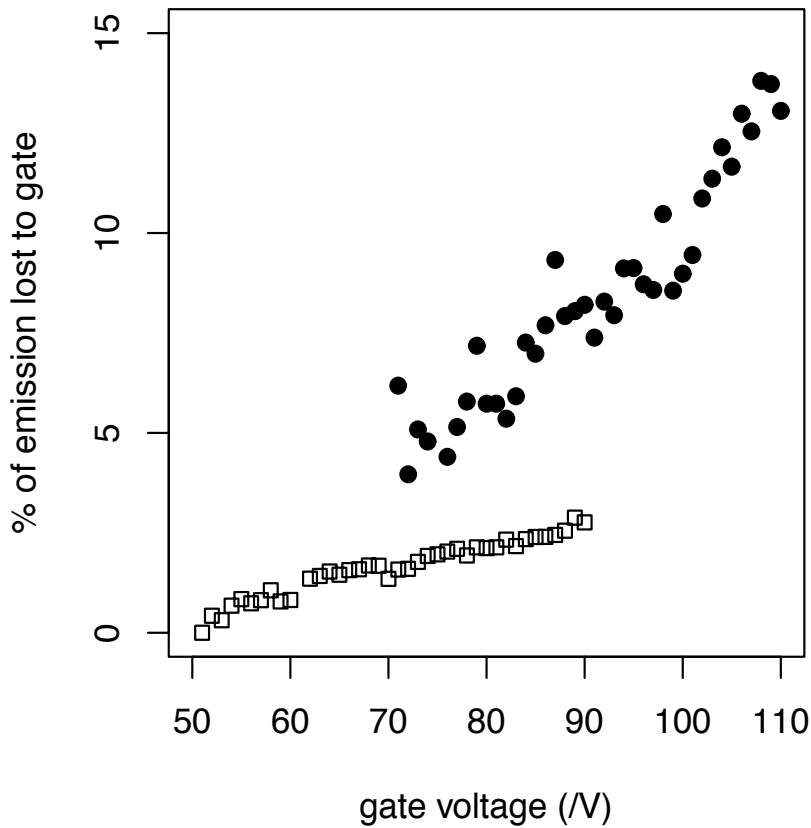

Figure 5

Figure 6

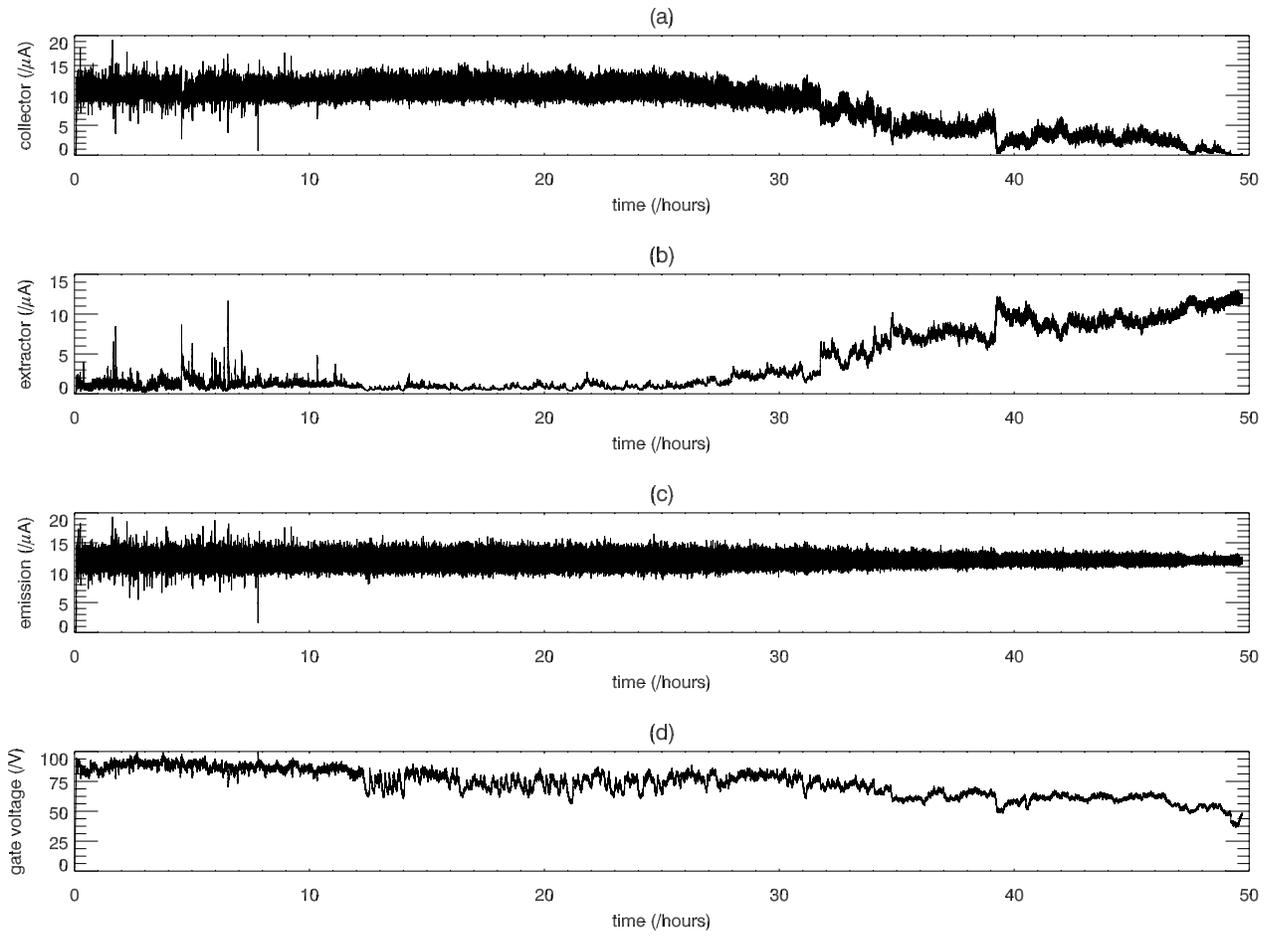

Figure 7

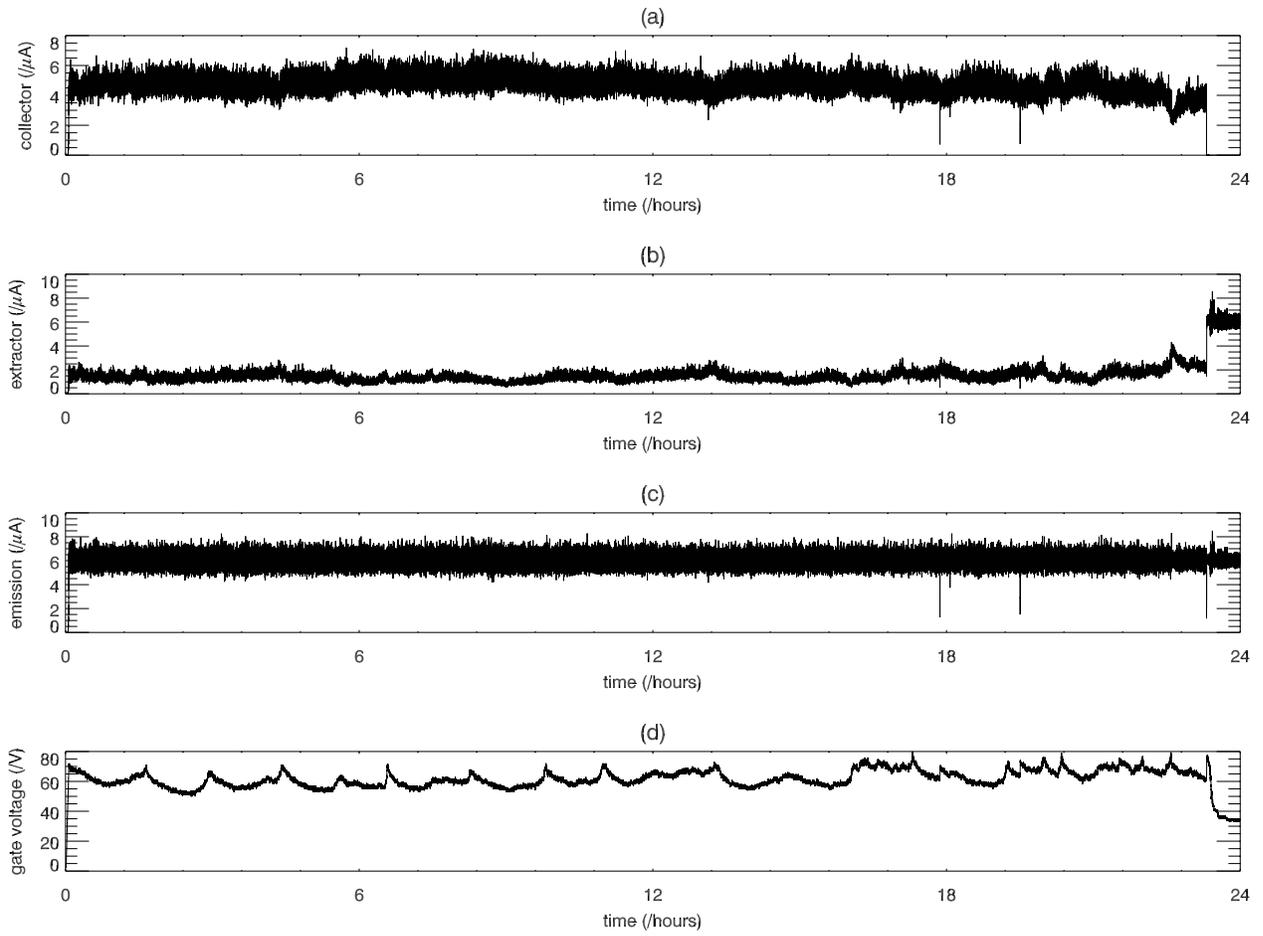

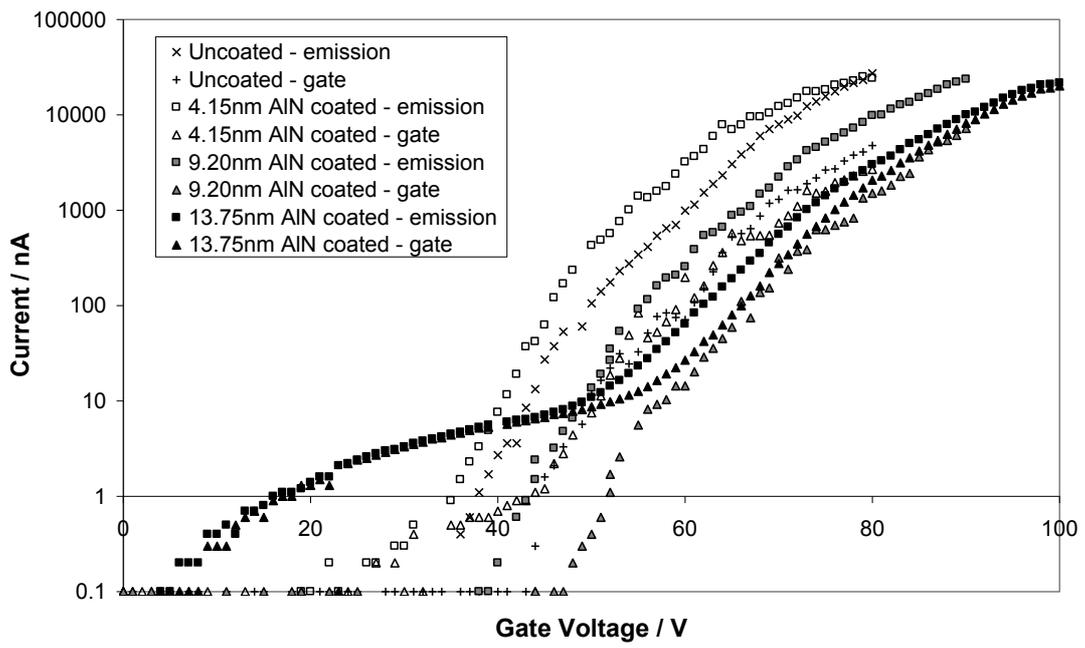

Figure 8